\documentclass[prb]{revtex4}

\usepackage{graphicx}

\begin{document}
\title{Fano fluctuations in superconducting nanowire single-photon detectors}
\author{A. G.Kozorezov$^1$, C. Lambert$^1$, F. Marsili$^2$, M. J. Stevens$^3$, V. B. Verma$^3$, J.P.Allmaras$^2$, M. D. Shaw$^2$, R. P. Mirin$^3$, Sae Woo Nam$^3$}
\address{$^1$Department of Physics, Lancaster University, Lancaster, UK, $^2$ Jet Propulsion Laboratory, California Institute of Technology, 4800 Oak Grove Dr., Pasadena, California 91109, USA, $^3$National Institute of Standards and Technology, 325 Broadway, Boulder, CO 80305, USA
}
\begin{abstract}
 Because of their universal nature, Fano fluctuations are expected to influence the response of superconducting nanowire single-photon detectors (SNSPDs). We predict that photon counting rate ($PCR$) as a function of bias current ($I_B$) in SNSPDs is described by an integral over a transverse coordinate-dependent  complementary error function. The latter describes smearing of local responses due to Fano fluctuations of the amount of energy deposited into electronic system.  The finite width, $\sigma$, of the $PCR$ vs $I_B$ arises from fluctuations in the energy partition between quasiparticles and phonons during the energy down-conversion cascade. In narrow-nanowire SNSPDs the local responses are uniform, and the effect of Fano-fluctuations on $\sigma$ is dominant. In wide-nanowire SNSPDs with strong coordinate dependence of local responses due to vortex-antivortex unbinding and vortex entry from edges, Fano-fluctuations smear singularities imprinted by vorticity on the transition part of $PCR$ curve. We demonstrate good agreement between theory and experiments for a series of bath temperatures and photon energies in narrow-wire WSi SNSPDs. The time-resolved hotspot relaxation curves predicted by Fano fluctuations  match the Lorentzian shapes observed in experiments over the whole range of bias currents investigated except for their tails .
\end{abstract}
\date{\today}
\maketitle\twocolumngrid

\section{Introduction}
    \quad The conversion of light into detectable excitations constitutes the key process in photodetection. Energy flow and relaxation pathways are the central topic which is equally interesting in many areas from traditional scintillators\cite{Williams} or new graphene-based materials\cite{Levitov1,Levitov2}. The efficiency of  detection depends on competition between different energy flow pathways.

    \quad Fano fluctuations describe variations in the number of charge carriers generated in a single-particle or single-photon sensor. Upon impact with a particle or absorption of a photon, energy is deposited in the sensor. This energy is partitioned between charged and neutral elementary excitations, for example between electrons and phonons. Fano fluctuations are caused by the branching processes and result in variations in the fraction of energy deposited in each system. Fano fluctuations are known to determine the theoretical limit of spectral resolution of many types of spectrometers, and are a limiting factor in the noise characteristics of CCDs and CMOS image sensors\cite{CMOS,Knoll}, as well as superconductor sensors such as superconducting tunnel junctions and microwave kinetic inductance detectors\cite{Kurakado, Rando}. Fano fluctuations may also be significant in sensors lacking an energy gap in the spectrum of elementary excitations, for example in superconducting transition edge microcalorimeters grown on a solid substrate\cite{Cabrera,Martin,Kozorezov2} and magnetic microcalorimeters\cite{Bandler,Miller}. The  relevant parameter of the detector material is the Fano factor, which quantifies the branching variance: a smaller factor indicates better resolving power.

    \quad To date, the role of Fano fluctuations in SNSPDs has not been discussed in the literature, likely because it was viewed as irrelevant. It has generally been assumed that an ideal SNSPD should exhibit sharp spectral and current thresholds for photodetection, characterized by a step function in detection efficiency when plotted as a function of bias current \cite{Semenov1}. The broadening of this step function into the sigmoidal shape observed in experiments has been attributed to inhomogeneities in nanowire width or thickness, or to variations in the position of the photon absorption site affecting vortex entry at the edge of the wire competing with unbinding of vortex-antivortex pairs away from the edge\cite{Semenov2,Engel,Bulaevskii1,Bulaevskii2,Vodolazov1,Vodolazov2,Renema,Semenov3,Vodolazov3}.

    \quad Here, we present a study of the influence of Fano fluctuations on the current and spectral dependence of the detection efficiency of SNSPDs. We found that a sigmoidal shape is expected even in the absence of these inhomogeneities. We show that the energy deposited into the electronic system by monochromatic photons fluctuates about the mean value due to the partition between quasiparticles (QP) and phonons. This occurs during energy down-conversion, with a variance  given by the Fano factor.  We show that photon counting rate vs. bias current in SNSPDs  in general is described by an integral over transverse coordinate-dependent  complementary error function with a width $\sigma$ determined by variance of Fano fluctuations.

    \section{Fano fluctuations in SNSPDs}

     \subsection{Energy down-conversion cascade}

\begin{figure}
  \centering
  \includegraphics[width=8cm]{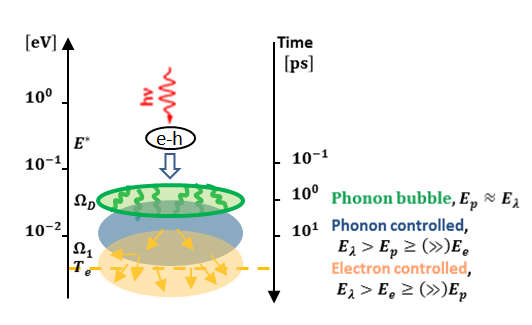}\\
  \caption{Schematic picture of photoelectron-hole energy down-conversion cascade in a metal}\label{fig:1}
\end{figure}

\begin{figure}
  \centering
  \includegraphics[width=8cm]{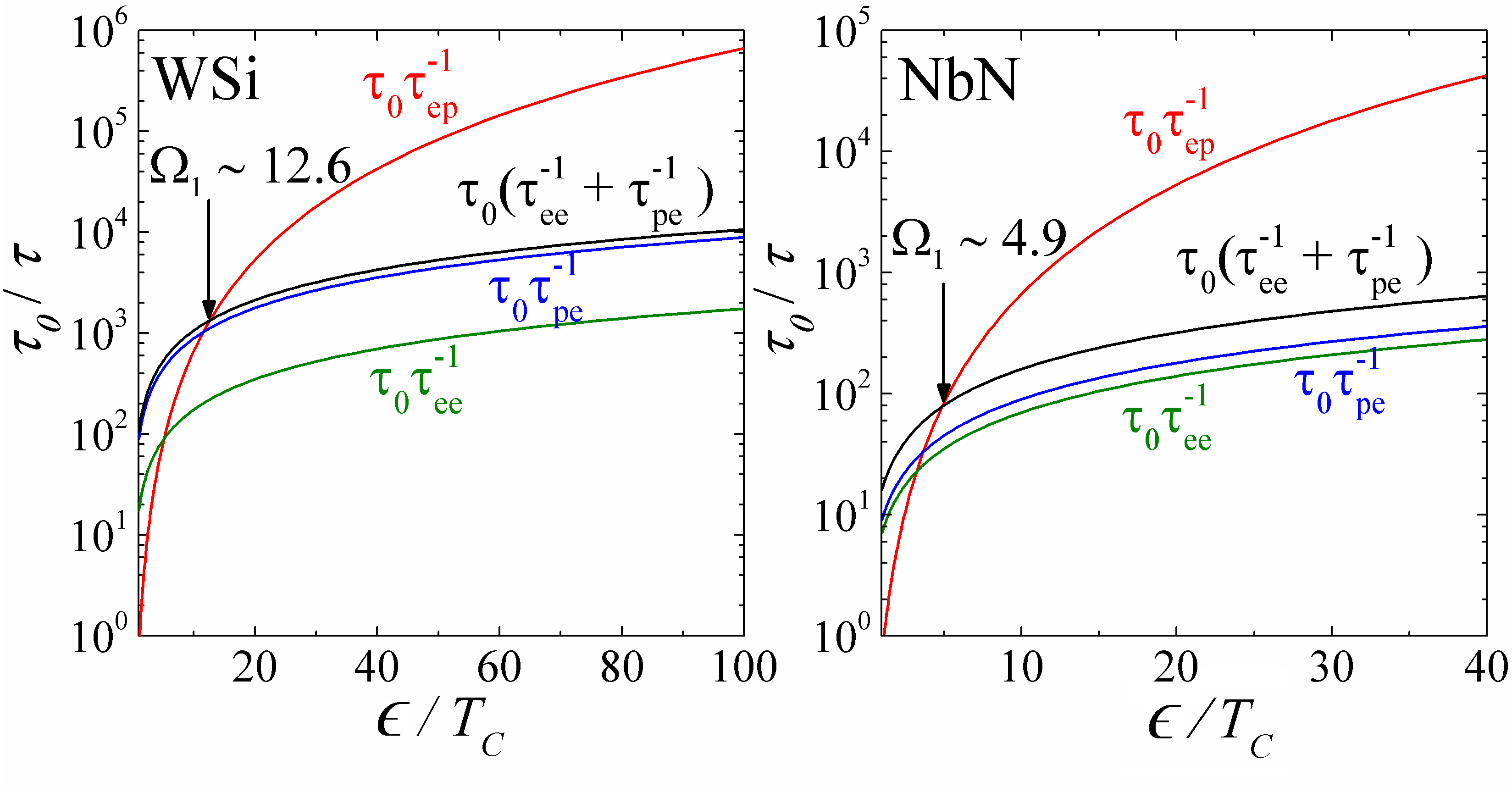}\\
  \caption{Electron-electron, $\tau_{ee}$, electron-phonon, $\tau_{ep}$, and phonon-electron, $\tau_{pe}$, scattering-out times vs excitation energy  in WSi and NbN. }\label{fig:3}
\end{figure}

    \quad We start by describing the first moments following the absorption of a photon in the SNSPD. We adapt the picture of energy down-conversion developed in our early work\cite{Kozorezov1} for a thin, disordered metal film. Fig.\ref{fig:1} schematically illustrates the evolving down-conversion cascade.
At $t=0$, a photon is absorbed and an electron-hole pair is generated. The sum of electron and hole energies is the energy of a photon, $E_\lambda$. On average half this energy is given to the electron and half to the hole. The energy and time scales in Fig.\ref{fig:1} are given  to illustrate an absorption of optical photon.  $E_1^*$ is the threshold energy separating the higher-energy interval, where electron-electron scattering with large momentum transfer dominates inelastic relaxation of the electron(hole) due to electron-phonon interaction. For excitations of energy $\epsilon<E_1^*$ the electron-phonon interactions totally dominate both the phonon-electron and electron-electron interactions until the mean energies of interacting electrons and phonons reach a lower threshold, $\Omega_1$, below which electron-electron and phonon-electron interactions in the  disordered film dominate. Usually $\Omega_1\ll\Omega_D$, where $\Omega_D$ is the Debye energy. Fig.\ref{fig:3} shows scattering-out relaxation times as a function of quasiparticle energy calculated from Fermi energy over the range of energies up to Debye energy.
Electron-electron scattering time as a function of energy was calculated from Altshuler and Aronov's formula for a disordered 2D normal metal\cite{Altshuler}, while electron-phonon and phonon-electron scattering times were taken from Chang\cite{Chang}.  It is important to note that the expressions for electron-phonon and phonon-electron times were taken neglecting vertex renormalisation by strong disorder, which is valid for the range of energies $\epsilon£l/\hbar£c\geq1$, where $l$ is the elastic mean free path for electrons and $c$ - is the sound velocity\cite{Schmid,Sergeev}. This inequality holds true for higher energies, while at lower energies, close to $\Omega_1$, the role of disorder becomes important. The characteristic time $\tau_0$ is taken for WSi and NbN to be 5-10 and 0.3-0.6 ns respectively, which falls into the expected range for the clean limit, consistent with the magnitude of the  Debye and critical temperatures if the median value of these is taken as the effective coupling strength\cite{Kaplan}. As follows from Fig.\ref{fig:3}, in both NbN and WSi the electron-phonon interaction dominates the electron-electron scattering rate over a significant spectral range. This means that the loss of high frequency non-equilibrium (athermal) phonons from the thin film, with thickness comparable to the phonon mean free paths, may be significant.

\quad Using the estimate for electron-electron scattering-out time for large momentum transfers for $\epsilon>\Omega_D$, we obtain that $\tau_{ee}(\epsilon)>\tau_{ep}(\epsilon>\Omega_D)=\tau_s$ for  $\epsilon<E_1$ with $E_1$ close to 1 eV\cite{Kozorezov1}. The energy loss for electron/hole of the initial pair due to electron-electron interactions with electrons of the Fermi distribution at equilibrium is slower. It can be roughly estimated from the expression for $\tau_{ee}^{out}(\epsilon)$\cite{Altshuler}
\begin{eqnarray}
&&\frac{1}{\tau_{ee}^{out}}(\epsilon)\propto\int_0^\epsilon\mathrm{d}\omega\int_0^\omega\mathrm{d}\epsilon'\int_0^\infty\mathrm{d}qq^2W_q\frac{1}{qV_F} \nonumber\\&&\dot{\epsilon}\propto\int_0^\epsilon\mathrm{d}\omega\omega\int_0^\omega\mathrm{d}\epsilon'\int_0^\infty\mathrm{d}qq^2W_q\frac{1}{qV_F}\nonumber\\&&
\dot{\epsilon}=-\frac{2}{3}\frac{\epsilon}{\tau_{ee}^{out}(\epsilon)}
\end{eqnarray}
where $\omega$ and $q$ are the energy and momentum transfer at the collision, $\epsilon'$ is the energy of a particle the electron collides with, $W_q$ is the matrix element describing the interaction, $V_F$ is Fermi velocity. Solving the last equation yields
\begin{eqnarray}\label{energy drop}
\frac{\epsilon(t)}{\epsilon_0}=\left[1+\frac{4}{3}\frac{t}{\tau_{ee}^{out}(\epsilon_0)}\right]^{-1/2}
\end{eqnarray}
where $\epsilon_0=\epsilon(0)$. The energy of an electron excited at $\epsilon$ as follows from (\ref{energy drop}) halves at $9/4\tau_{ee}^{out}(\epsilon)$. The loss of half of the energy due to sequential emission of phonons takes $\sim(\epsilon/2\Omega_D)\tau_s$. Both rates are equal at $E_1^*$, $\displaystyle{E_1^*=E_1\left(\frac{9}{2}\frac{\Omega_D}{E_1}\right)^{1/3}}$. For $\Omega_1<\epsilon<E_1^*$ the electron-phonon scattering is the dominant mechanism of electron/hole energy relaxation.
With $E_1\sim1$ eV and $\Omega_D\sim30$
meV we obtain for typical materials $E_1^*\sim$ 500 meV.  When the energy of initial electron/hole (or secondary excitations) is close to $E_1^*$ their subsequent cooling proceeds mostly via sequential emission of phonons. Since the characteristic values for $\tau_s$ are tens of fs, unless the materials have a small Debye energy, cooling from 500 meV down to $\Omega_D$ will last a fraction of a picosecond. By the end of it, which completes before the lifespan of the first emitted phonons, most of the photon energy has been transferred to high energy (Debye) phonons. It is convenient to consider this highly non-equilibrium state as the natural initial condition for the subsequent evolution of interacting quasiparticles and phonons. In Fig.\ref{fig:1} this condition is called a phonon bubble.

\subsection{Modelling Fano-fluctuation}

    \quad We consider a simple model of photon detection\cite{Marsili1,AGK}. Following the absorption of a photon of energy $E_\lambda$, a fraction of the energy, $E<E_\lambda$, is deposited into electronic excitations. During the energy down-conversion cascade, a hotspot is created\cite{Kozorezov1}. For a narrow and thin wire we assumed that the hotspot spans the width of the wire, $W$, occupying volume $V_{HS}=WL_{HS}d$,  with $L_{HS}\geq�W$ being its length along the wire and $d$ is the wire thickness. In this situation $L_{HS}$ can be determined in tomography experiments\cite{Marsili1}. We assume that after completion of the cascade the quasiparticle distribution is thermalized owing to a strong electron-electron interaction in a disordered nanowire. Therefore, we assumed that after the absorption of a photon the QP temperature instantly increased from bath temperature, $T_b$, to excitation temperature, $T_{ex}$. The hotspot excitation temperature is determined from the thermal balance
    \begin{eqnarray}\label{thermal balance}
    E_{HS}(I_B,T_{ex},B)-E_{HS}(I_B,T_b,B)=\int_{T_b}^{T_{ex}}\mathrm{d}T'~~~~~~\nonumber\\C(T',I_B,B)=E^{QP}_{HS}(I_B,T_{ex},B)-E^{QP}_{HS}(I_B,T_b,B)
    +\nonumber\\2N(0)V_{HS}k_B\int_{-\infty}^\infty�\mathrm{d}\xi\int_{T_{ex}^{-1}}^{T_{b}^{-1}}\mathrm{d}\beta£\frac{1}{\exp(\beta\epsilon)+1}\frac{\partial£\epsilon}{\partial\beta}=E~~~
    \end{eqnarray}
    where  $E_{HS}(I_B,T,B)$ is the internal energy of the hotspot, $N(0)$ is the normal state density of states at the Fermi level, $\epsilon=\sqrt{\xi^2+\Delta^2(I_B,T,B)}$ is the quasiparticle energy, $\Delta$ is the order parameter in current carrying nanowire, $B$ is the external magnetic field and $E$ the actual amount of energy deposited in the electronic system.
\begin{eqnarray}\label{QP energy}
E_{HS}^{QP}(I_B,T,B)=4N(0)V_{HS}\int_0^\infty�\mathrm{d}\epsilon\frac{\epsilon\rho(\epsilon,I_B,T,B)}{\exp\left(\epsilon/T\right)+1}
\end{eqnarray}
Here $\rho(\epsilon,T,I_B,B)$ is the density of states within the hotspot in units of $N(0)$. The dependencies on bias current $I_B$, temperature $T$ and magnetic field $B$ originate from the pair-breaking energy and the order parameter being functions of $T$, $I_B$ and $B$. The last term in the expression (\ref{thermal balance}) originates from the dependence of QP dispersion relations in the hotspot on temperature.

\quad In an ideal SNSPD there is a count event every time the energy $E$ exceeds the threshold $E^*$, determined from
 \begin{eqnarray}\label{threshold}
    E_{HS}(I_B,T_b,B)+E^*=E_{HS}(I_B,T_s,B)
    \end{eqnarray}
where $T_s$ is the temperature at which the hotspot undergoes transition from the superconducting to the normal state. This temperature depends on bias current and external magnetic field and can be found for a narrow wire using the solution of the Usadel equation for density of states\cite{A. Anthore} and the dependence of the order parameter on current, temperature and magnetic field\cite{Marsili1,AGK}.

 \quad In wider wires the characteristic shape of a hotspot is more complicated. If the photon absorption site is not close to the edge of the wire, then the hotspot is cylindrical. Its radius, $R_{HS}$, is smaller than
its width, $W >R_{HS}\gg£d$, the hotspot does not span the wire width, and its creation results in a current density redistribution. The latter
depends on the lateral coordinate of the absorption site, $\textbf{r}$ (2-dimensional geometry). Finally, in
very thick wires the hotspot may be spherical (or close to semi-spherical for shallow absorption), with the radius $R_{HS} \ll£\min\{d,W\}$. The geometry of current flow in this wide and thick nanowire becomes 3-dimensional.

\ In wide thin nanowires, the change in the current flow  facilitates vortex entry at the edges of the nanowire or unbinding of vortex-antivortex pairs. In this situation, Fano fluctuations result in supercurrent density fluctuations and their description becomes more complicated. The fluctuating order parameter, current density and temperature in the hotspot become dependent on the coordinate of the absorption site and are connected through a more complicated relation. The coordinate-dependent detection current, $I_{det}(y,E,T_b,B,\lambda)$ is introduced through the appropriate simulation\cite{Engel,Vodolazov1}, and the implicit condition for the $E^*$ threshold to trigger the detection click can be written as
\begin{eqnarray}\label{WW condition}
I_B=I_{det}(y,T^*,T_b,B,\lambda)=I_{det}(y,E^*,T_b,B,\lambda)
\end{eqnarray}
where $T^*=T(E^*)$ is the threshold temperature of the hotspot corresponding to energy deposition $E^*$.

 \quad The normalised  probability distribution $P(E)$ describing energy deposition into the electronic system is Gaussian, $\int_{-\infty}^\infty\mathrm{d}EP(E)=1$,
\begin{eqnarray}\label{s lineshape}
P(E)=\frac{1}{\sqrt{2\pi}\sigma}\exp\left[-\frac{(E-\bar{E})^2}{2\sigma^2}\right]
\end{eqnarray}
The distribution is centered around a mean value $\bar{E}=\bar{\chi}�E_\lambda$, where $\bar{\chi}<1$, $\chi = E /E_\lambda$ is the photon yield, which we defined as the ratio of the energy deposited in the hotspot after the absorption of the photon ($E$) to the photon energy ($E_\lambda$).  The full width at half maximum of the distribution $P(E)$ is $2\sqrt{2\ln2}\sigma$, where $\sigma$ is the  variance. Since a count occurs when $E\geq�E^*$, simple integration yields the normalised $PCR$ for a narrow wire in the form
\begin{eqnarray}\label{PCR}
&&PCR^{NW}=\int_{E^*}^{\infty}\mathrm{d}EP(E)=\frac{1}{2}\mathrm{erfc}\left(\frac{E^*-\bar{E}}{\sqrt{2}\sigma}\right)\nonumber\\&&
=\frac{1}{2}\mathrm{erfc}\left[\frac{E(I_B,T_s,B)-E(I_B,T_b,B)-\bar{E}}{\sqrt{2}\sigma}\right]
\end{eqnarray}
For a wide wire the photon count occurs if the bias current exceeds the minimum of detection current, which is a function of the coordinate $y$ across the wire and depends on hotspot temperature (deposited energy), bath temperature, magnetic field, photon wavelength, $I_{det}\left(y,T(E),T_b,B,\lambda\right)=I_{det}\left(y,E,T_b,B,\lambda\right)$. In this situation
\begin{eqnarray}\label{PCRWide}
&PCR^{WW}=\int_{0}^{\infty}\mathrm{d}EP(E)\displaystyle{\frac{1}{W}}\int_{-W/2}^{W/2}\mathrm{d}y\Theta\left[
I_B-\right.\nonumber\\&\left.I_{det}\left(y,E,T_b,B,\lambda\right)\right]=\int_{0}^{\infty}\mathrm{d}EP(E)w\left(E,T_b,B,\lambda\right)=\nonumber\\&
\displaystyle{\frac{1}{W}}\int_{-W/2}^{W/2}\mathrm{d}y~\mathrm{erfc}\left[\displaystyle{\frac{I_B-I_{det}\left(y,\chi£E_\lambda,T_b,B,\lambda\right)}
{\sqrt{2}\sigma£I'_{det}\left(y,\chi£E_\lambda,T_b,B,\lambda\right)}}\right]
\end{eqnarray}
where $\Theta(x)$ is the Heaviside function, $w\left(T(E),T_b,B,\lambda\right)$ is the fraction of the wire width where generation of a vortex results in formation of a normal domain across the wire, and $I'_{det}\left(y,\chi£E_\lambda,T_b,B,\lambda\right)=\partial£I'_{det}\left(y,E,T_b,B,\lambda\right)/\partial£E£|_{E=\bar{\chi}£E_\lambda}$
In wide nanowires, the expression (\ref{PCRWide}) must be further convolved if the light polarization results in a  spatial profile of the absorption sites.

\quad The variance is determined from two statistically independent processes as $\sigma^2=\sigma_1^2+\sigma_2^2$. Following photon
absorption, a rapid process of energy down-conversion is
initiated, engaging numerous electronic and phonon excitations.
Phonons can be divided into two groups. Non-pair-breaking phonons with energy $\hbar\Omega$ smaller than twice the order parameter $2\Delta$ are decoupled from the condensate and can only be reabsorbed by excited QPs. In
thin films we neglect their reabsorption,
assuming the re-absorption time to be considerably longer than their escape into a substrate or thermalization time. In contrast,
higher energy phonons can break Cooper pairs and exchange energy with the electronic system.
In a bulk superconductor, if the distance from the absorption site to escape interface far exceeds the phonon mean free path, none of the pair-braking phonons escapes into the thermal bath before QPs thermalise.  By  the end of  down-conversion the energy of a photon
splits between QPs and non-pair-breaking phonons. $\sigma_1$ describes
statistical fluctuations of $E$ originating from fluctuations
of numbers of pair-breaking phonons under the assumption
that none of the pair-breaking phonons from the down-conversion cascade escapes. It can be written $\sigma_1^2=F\varepsilon£E$, where 2$\varepsilon$ is the mean energy which is
necessary to spend in order to generate one pair of QPs. In superconducting films with thickness comparable
to the mean free path of pair-breaking phonons,
some of the highly energetic (athermal) phonons will escape.
Thus, instead of $E$ we must write $E'=(1-\chi_a)E$,
where we introduce $0 < \chi_a < 1$ to account for the average
fraction of athermal phonons escaping from the
nanowire prior to random partition of energy between QPs and
non-pair-breaking phonons, which are de-coupled from condensate. The Fano factor in most
superconductors is $F\simeq 0.2$\cite{Kurakado,Rando}.

\quad The energy loss from the film due to escaping athermal
phonons is a random process, depending on mean
free paths of phonons and transmission characteristics
into the substrate. Therefore, there appears the second
independent contribution to the variance, $\sigma_2$. If all
athermal phonons are re-absorbed in the film, which
is the case for thick films illuminated from the top, then $\chi_a=0$ and $\sigma_2=0$ while
$\sigma_1^2 = F\varepsilon£E$ becomes the known Fano fluctuations variance. $\sigma_2$
originates from fluctuations in the transmission of
athermal phonons to the substrate\cite{Martin,Kozorezov2}:
$\sigma_2^2 = J(E_\lambda)\varepsilon£E_\lambda$, we call this contribution
phonon down-conversion noise. This contribution to Fano noise has also been expressed in conventional form\cite{Kozorezov4} with the factor $J(E_\lambda)$.
The accurate determination of Fano-factor $J(E_\lambda)$ is not an easy problem. It involves evaluation of statistical fluctuations of loss of successive generations of athermal phonons, which are part of evolving distribution of the down-conversion process. We may, however, derive the lower limit to $J(E_\lambda)$. While lower-energy longer-living phonons from later generations account for the majority of energy loss, the opposite is true for fluctuations of energy loss. With multiplication of phonon numbers and decrease of their mean energy, the relative fluctuation of energy loss from phonons from later-generations decreases. We estimate the contribution to $J(E_\lambda)$ from phonons of first generation (phonon bubble) designating it $J^1(E_\lambda)$ and ignore contributions from subsequent athermal phonon generations arriving in the lower limit, $J^1(E_\lambda)<J(E_\lambda)$.

\quad In a thin film, photons are absorbed homogeneously through the depth of the film, and, similarly, phonons of the first generation (phonon bubble) are also homogeneously generated while the energetic photoelectron(hole) performs a random walk in the film. Thus,  taking the limit $d\rightarrow0$ and $m=0$ we simplify the expression for $J^1(E_\lambda)$, which was derived in earlier work\cite{Kozorezov4} and obtain
\begin{eqnarray}\label{Fano J}
&\displaystyle{J^1=2\frac{\Omega_D}{\varepsilon}\frac{l_{pe,D}}{d}\int_{\cos(\theta_c)}^1\mathrm{d}\xi\xi\eta(\xi)\left\{\frac{1}{4}
\left(\frac{l_{pe,D}}{d}\right)^4-\right.}\nonumber\\&\displaystyle{\left.\int_0^{d/l_{pe,D}\xi}\mathrm{d}xx^3\left[e^{-x}+\frac{1}{2}\eta(\xi)\sin^2(\theta_c/2)\left(1-e^{-2x}\right)\right]\right\}}\nonumber\\
\end{eqnarray}
Here $\theta_c$ is the angle of total internal reflection, so that phonons that impinge the escape interface at larger angles stay inside the plane parallel film until they undergo scattering-assisted conversion and move at smaller angles $\theta<\theta_c$, $\eta$ is the phonon transmission coefficient through the interface with the substrate for incidence below the critical angle, and $l_{pe,D}$ is the mean free path of Debye phonons with respect to absorption by electrons. This result has the following meaning. The variance of the number of phonons emitted into the critical cone from any energy interval follows a binomial distribution. The contributions from phonons of different energies are statistically independent. Therefore, total variance is the sum (integral) of contributions from individual groups of phonons of the flat distribution of phonons of the first generation. Individual factors in the expression (\ref{Fano J}) reflect : $\Omega_D$ - the dominant phonon energy in the phonon bubble, $\frac{l_{pe,D}}{d}$ - probability of survival until reaching the escape interface, the inner integral - averaging phonon contributions over their distribution, accounting for their probabilities to reach the interface depending on energy and angle of incidence within the critical cone. Note that a binomial distribution is close to a normal distribution in expression (\ref{s lineshape}) under the assumption of large phonon numbers in each energy interval. Thus, integration over the spectrum works well for phonons of later generations. For first-generation phonons it works for the dominant phonon group with energies close to Debye energy, and the second integral in (\ref{Fano J}) is an approximation of the dominant phonon contribution.

\quad The total variance due to Fano noise can be written as $\sigma^2=F_{eff}\varepsilon�E_\lambda$, where the effective Fano-factor is  $F_{eff}=F(1-\chi_a)+J$. Our estimate of $J$ relies on using the simplest model of one phonon-mode with a  linear dispersion relation (Debye) approximation. For phonons of the first few generations we have $\epsilon£l/\hbar£c\geq1$ , and our
estimate of $J^1$ is not affected by disorder. Given these approximations this estimate provides only a rough indication of the magnitude of the effect.

\quad In the next section we compare our simulation with experiments on WSi SNSPDs. Because of the large acoustic mismatch between WSi (W$_3$Si) and the SiO$_2$ substrate, even for normal incidence we have $\chi\leq0.5$. Taking $\chi=0.5$ for WSi on amoprphous SiO$_2$ substrate we obtain $1.0\leq£J\leq1.5$ for $5\leq\tau_0\leq10$ ns. For NbN on SiO$_2$, a similar calculation yields $0.8\leq£J\leq1.0$ for $0.5\leq\tau_0\leq1.0$ ns.
With these estimates we conclude that phonon down-conversion noise even due to phonons of the first generation dominates Fano fluctuations,  $\sigma_2\gg\sigma_1$. This is not surprising because the fluctuations of the number of phonons falling within the escape cone and the fluctuations of  ultimate number of quasiparticles are both proportional to the square root of the respective mean numbers. Fluctuations for athermal phonons are strong, partly because of their large energy and hence smaller numbers, but more importantly because only a significantly smaller fraction of all athermal phonons (those within the escape cone) contribute to fluctuations in the corresponding energy loss. The subsequent phonon generations also contribute to fluctuations increasing $\sigma_2$, although each subsequent contribution becomes smaller, because of the twofold increase of the number of phonons with every generation. Thus, even if we overestimated $J^1$, there is still an extra contribution due to lower energy phonons, so that most probably $\sigma_2\gg\sigma_1$.  Because we cannot provide credible estimates for contributions to overall variance from later phonon generations we will use $F_{eff}$ as a fitting parameter. The likely range of variation of $F_{eff}$ must be consistent with our estimate for $J^1$.

\subsection{The effect of Fano-fluctuations on the shape of $PCR$ curves in narrow, homogeneous SNSPDs}

\quad The shape of $PCR$ curve vs bias current at fixed values of external parameters $T,~\lambda,~B$ is affected by the bias dependence of the energy terms in the numerator of the argument of the complimentary error function in equation (\ref{PCR}), i.e $E(I_B,T_s,B)-E(I_B,T_b,B)-\bar{E}$. Also in the current-carrying superconductor both variances $\sigma_1$ and $\sigma_2$ depend on $I_B$. $\sigma_1$ depends on the  order parameter at critical point $\Delta_s$, while $\sigma_2$ due to contributions of phonons of later generations may weakly depend on the threshold energy $\Omega_1$, because both $\Delta_s$ and $\Omega_1$ depend on $I_B$. We defined the cutoff current of the SNSPD at fixed wavelength, bath temperature, and magnetic field, $I_{co}(\lambda,T_b,B)$, as the inflection point of the $PCR$ vs $I_B$ curve and used the approximation  $E(I_B,T_s,B)-E(I_B,T_b,B)-\bar{E}\approx\alpha^{-1}(I_B,\lambda,T_b,B)(I_B-I_{co}(\lambda,T_b,B))$ to obtain
\begin{eqnarray}\label{PCRIB}
&&PCR=\frac{1}{2}\mathrm{erfc}\left[\frac{I_B-I_{co}(\lambda,T_b,B)}{\sqrt{2}\sigma(I_B,T_b,B)\alpha(I_B,\lambda,T_b,B)}\right]~~~~
\end{eqnarray}
This is the modification of the equation (\ref{PCRWide}) for the case of a narrow wire with a detection current, that  is not dependent on $y$-coordinate across the wire. As seen from (\ref{PCRIB}), $PCR$ as a function of bias current differs from the ideal complimentary error function because of the current dependence of $\alpha$ and $\sigma$ in the denominator. If the width of the $PCR$ curve is small relative to de-paring current, we may keep only the first term in series expansion of $E(I_B,T_s,B)-E(I_B,T_b,B)-\bar{E}$. This corresponds to $\alpha$ being independent of bias current. Similarly, if the width of the $PCR$ curve is small relative to the de-paring current, the dependence of $\sigma$ on bias current can also be neglected. Therefore, deviations of $PCR$ from an ideal complimentary error function reflect either the specific features of Fano-fluctuations in the current-carrying superconducting nanowire (through $\sigma$) or strong non-linearity of the SNSPD response (through $\alpha$).

\quad The down-conversion in a strongly disordered current-carrying nanowire is different from typical superconducting sensors with no significant disorder, such as superconducting tunnel junctions and many types of microwave kinetic inductance detectors. In these detectors, the temperature of the hotspot does not rise close to the critical temperature $T_C$ even for photon energies as high as a few keV. The transport current is small compared to the de-pairing current and the order parameter stays close to its zero current-temperature value. Therefore, the energy required to generate a quasiparticle is constant during the downconversion process, $\varepsilon=1.75\Delta(0)$ . By contrast, in an SNSPD at the edge of the transition to the normal state, the temperature of the hotspot reaches the critical value $T_s$, while the order parameter remains nonzero at the level $\Delta(I_B,T_s)$, but is significantly smaller than its zero temperature value for a given bias current. As a result, some of the non-pair-breaking phonons with energies below $2\Delta(t)$ that were emitted shortly after photon absorption regain their pair-breaking capability later, when their energy starts exceeding the decreasing $2\Delta(t)$. This happens if down-conversion and thermalisation times are faster than phonon escape time from the film. In this asymptotic limit $\varepsilon=1.75\Delta(T_s)$, and it decreases with bias current, while $T_s$ increases\cite{AGK, Marsili1}. The corresponding $\sigma_1$ also decreases, resulting in a steeper rise of the $PCR$ vs $I_B$. The dependence of the second contribution $\sigma_2$ on $I_B$ is weaker,  reflecting the fact that $\sigma_2$ is the contribution from the initial generations of excitations when the order parameter is much less affected. Therefore, to experimentally observe the bias dependence of the variance $\sigma$, $\sigma_1$ and $\sigma_2$ must be comparable in magnitude. This situation is easier to fulfil in larger gap NbN than in WSi.

\begin{figure}
  \centering
  \includegraphics[width=7cm]{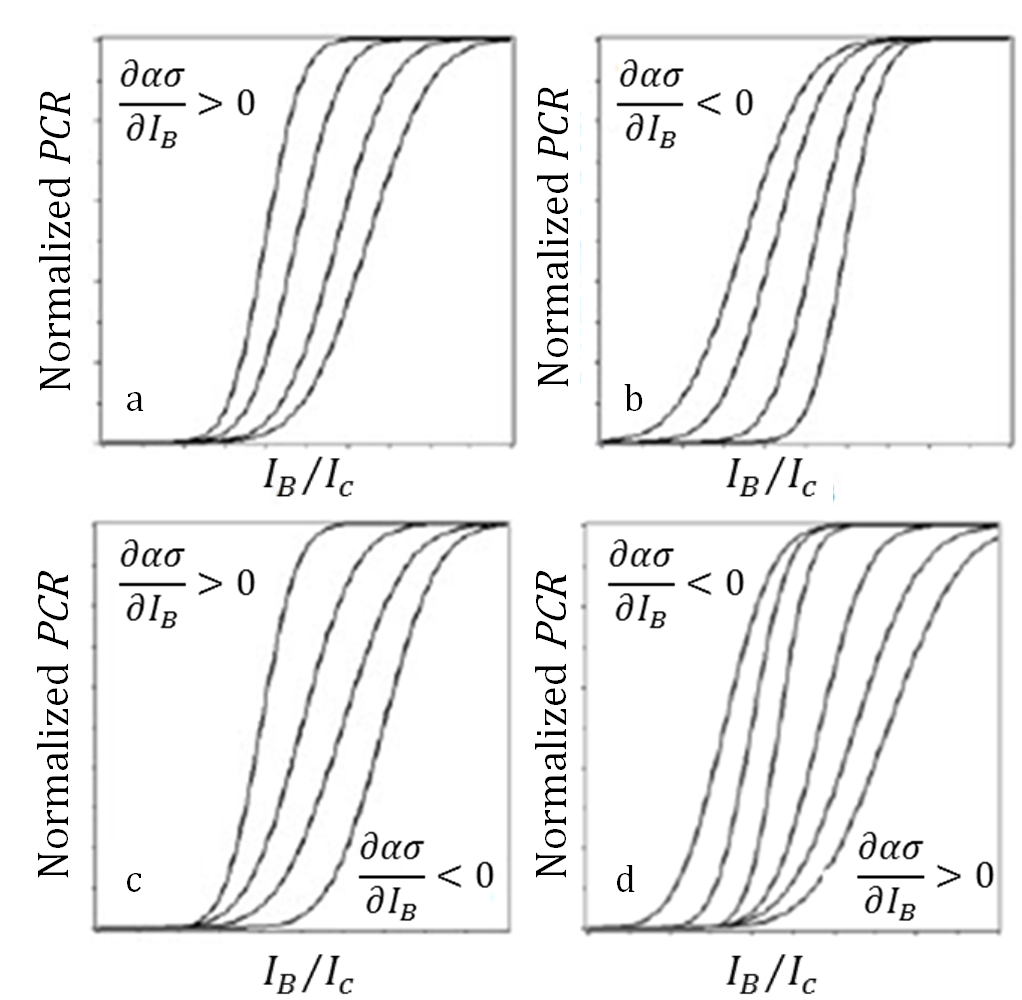}\\
  \caption{Predicted patterns of $PCR$ vs $I_B$}\label{fig:4}
\end{figure}

\quad The other factor, $\alpha$, depends on bias current considerably stronger, because it is determined by a highly non-linear function of bias current $E(I_B,T_s,B)-E(I_B,T_b,B)-\bar{E}$,  $\displaystyle{\alpha^{-1}(I_B,\lambda,T_b,B)=\frac{\partial\left(E(I_B,T_s,B)-E(I_B,T_b,B)\right)}{\partial£I_B}\Big{|_{\bar{\chi}£E_\lambda}}}$ (neglecting potential dependence of $\bar{E}$ of $I_B$).
It can be analysed only numerically. With the factor $\sigma(I_B,T_b,B)\alpha(I_B,\lambda,T_b,B)$ being independent of bias current, or if estimating it at inflection
points of experimental curves is a good approximation, then the predicted $PCR$ curves with any other parameters ($T_b,~\lambda,~ B$) changing must exhibit parallel
shifts along $I_B$ axis. With a stronger non-linear dependence of the factor $\sigma(I_B,T_b,B)\alpha(I_B,\lambda,T_b,B)$ on $I_B$, the predicted $PCR$ curves will exhibit changes of slopes of the
rising parts with onsets at different bias currents. Moreover, in general $\sigma(I_B,T_b,B)\alpha(I_B,\lambda,T_b,B)$ is non-monotonic in the
transition range. The four speculative patterns of $PCR$ vs $I_B$ can therefore be predicted as shown in Fig.\ref{fig:4}. While  the pattern in Fig.\ref{fig:4}a is not uncommon,
the pattern resembling Fig.\ref{fig:4}d was observed experimentally in narrow, $W$=30 nm, NbN SNSPDs\cite{Marsili2}.

\subsection{The effect of Fano-fluctuations on the shape of $PCR$ curves in wide SNSPDs}

\begin{figure}
\centering
  \includegraphics[width=7cm]{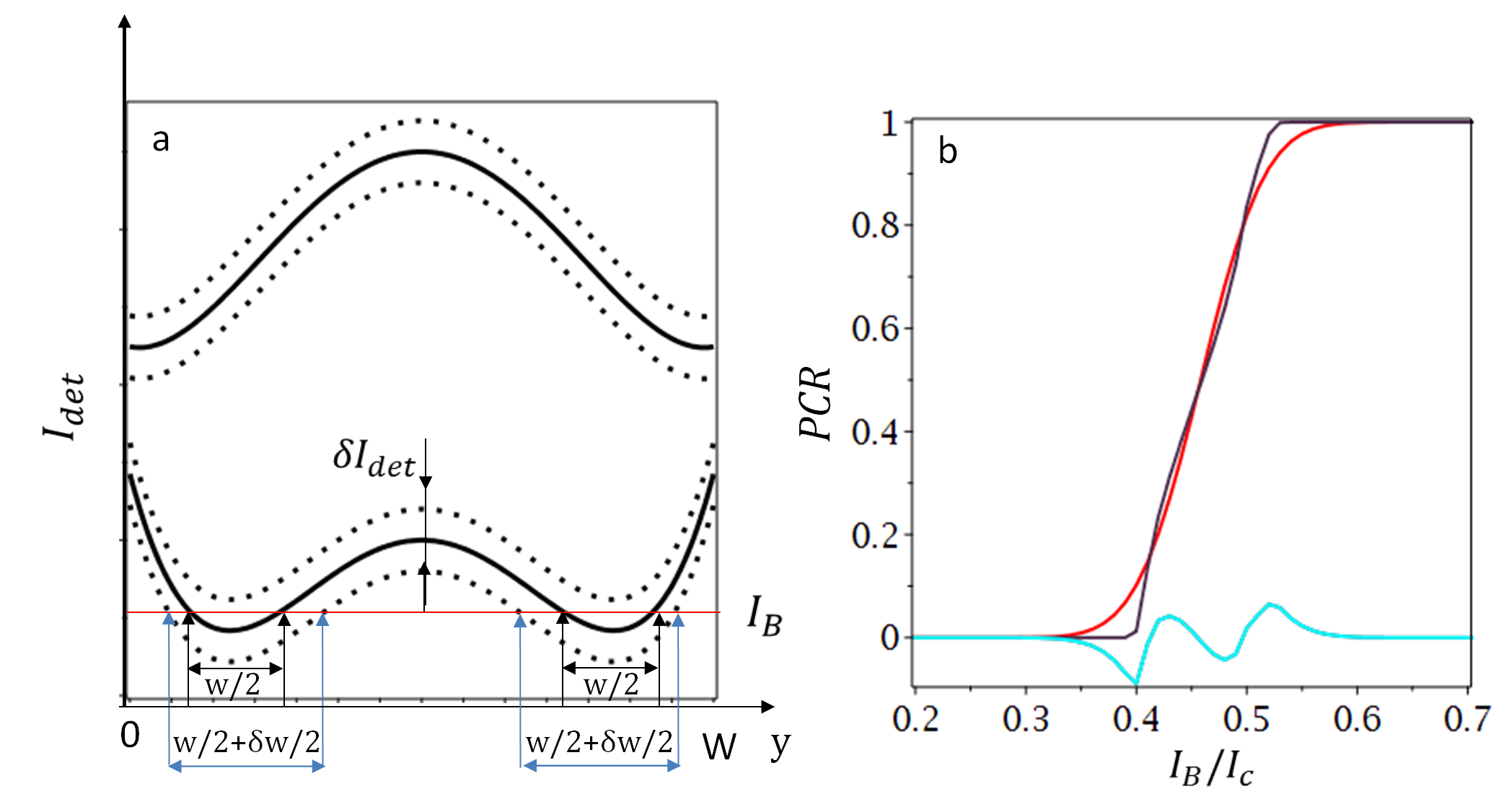}\\
  \caption{a. Schematic profiles of $I_{det}$ vs transverse coordinate $y$: top curve - bell shaped, bottom - w-shaped. b. $PCR$ for w-shaped profile of $I_{det}$, best fit error function shape -red and difference between the two shapes -cyan  }\label{fig:5}
\end{figure}

\quad In wide SNSPDs, the sensor response depends on the $y$-coordinate of the absorption site across the nanowire. In a spatially homogeneous nanowire this dependence will affect the
shape of the $PCR$ curves in the transition region, reflecting the curvature of the predicted dependence of detection current vs coordinate $y$. Two shapes of
$I_{det}(y,E,T_b,\lambda)$ are discussed in the literature: bell-shaped \cite{Engel} and w-shaped\cite{Vodolazov1,Vodolazov4}. Fig.\ref{fig:5}a shows schematically these $I_{det}$ profiles.
For illustration we will take the w-shaped profile depicted in Fig.\ref{fig:5}a with averaged detection current at one half and a difference between maximum and
minimum detection currents of $\sim$0.1 of the critical depairing current.  We obtain the $PCR$-curve shown in Fig.\ref{fig:5}b. Fano-fluctuations will smear this curve. However, in
order to evaluate this smearing, apart from the variance $\sigma$ we must evaluate the sensitivity of
$I_{det}(y)$ relative to fluctuations of the deposited energy, the derivative in the denominator in  (\ref{PCRWide}). This depends on the material properties and detection model.
Instead, shown superimposed is the closest fit error function curve and difference between the two curves. As is seen, the difference between the two curves can be resolved
experimentally allowing us to study the co-ordinate dependent response, on the background
of either normally distributed static spatial inhomogeneities or dynamic Fano-fluctuations.  Similarly, the distinguishing features of the $PCR$ curve for the bell-shaped profile of detection current can also
be resolved experimentally.

\quad Observation of smooth error function-like sigmoidal shapes in wide-wire SNSPDs will indicate a much weaker co-ordinate dependence than shown in Fig.\ref{fig:5}a. Such a weak co-ordinate dependence on its own cannot be responsible for the observable width of the transition region, $\delta£I_B$,  $I_{det,max}-I_{det,min}\ll\delta£I_B$.  In this case the shapes of $PCR$ curves of wide-wire  SNSPDs will closely resemble $PCR$ curves of narrow-wire  SNSPDs. Spatial  ingomogeneity connected to vortex initiation is totally smeared by fluctuations, which also determine the width of the  transition region, $\delta£I_B$. Moreover, hot belt and hot spot models\cite{Vodolazov4} will further merge with appropriate replacements $T_s\leftrightarrow£T^*$, $I_{co}\leftrightarrow£I_{det}$ and $\displaystyle{\alpha(I_B,\lambda,T_b,B)}\leftrightarrow£I'_{det}\left(y,\chi£E_\lambda,T_b,B,\lambda\right)$.

\section{Comparison with experiment}

\subsection{$PCR$ vs $I_B$ curves}
\begin{figure}
  \centering
  \includegraphics[width=6cm]{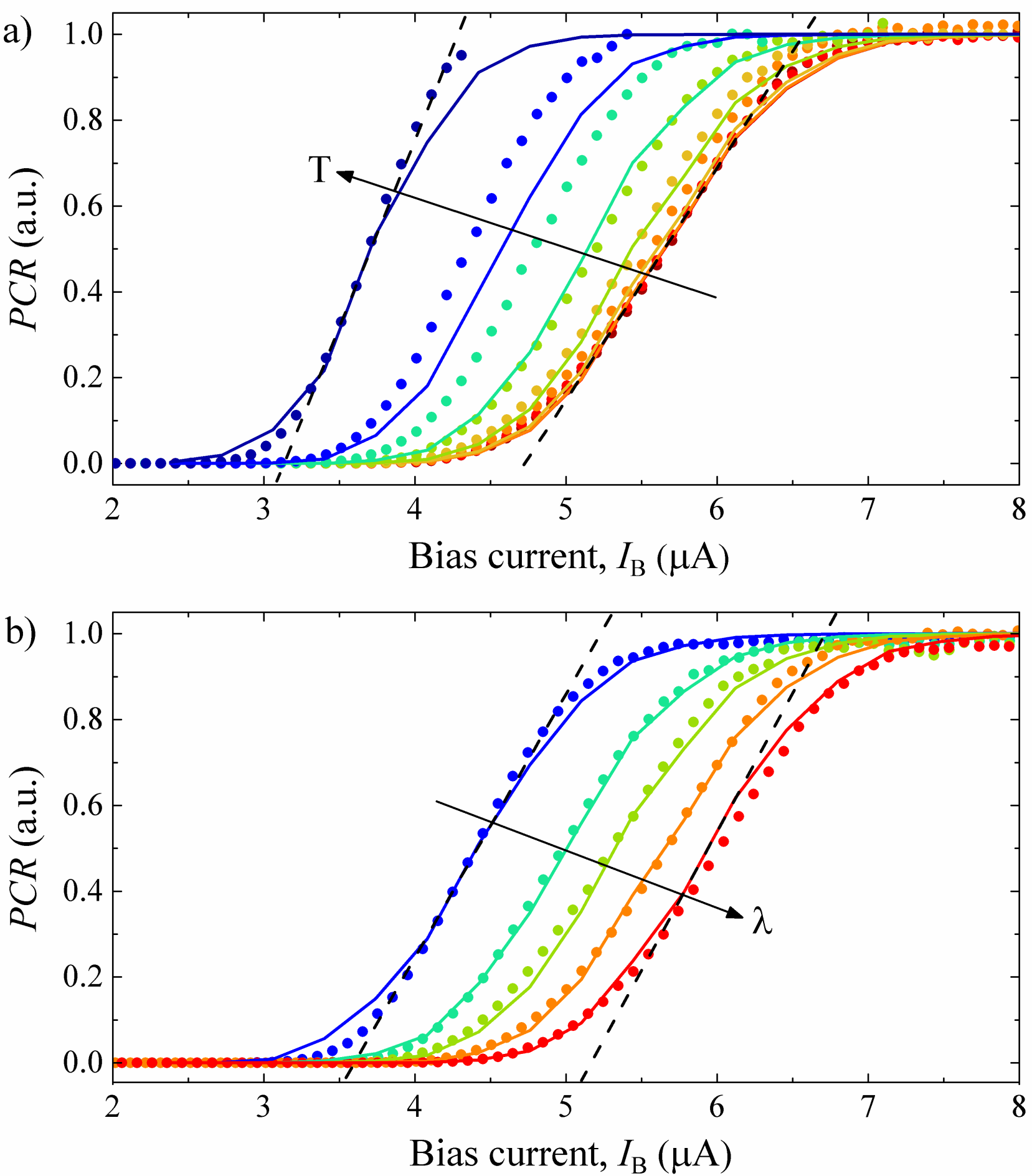}\\
  \caption{Photon count rate($PCR$) for a WSi SNSPD operated in the single-photon detection regime as a function of bath temperature a) from 250 mK to 2 K with the increment 250 mK and wavelength b) $\lambda$=1200, 1350, 1450, 1550 and 1650 nm. Solid curves-theoretical simulation, solid circles - experiment.  Dashed lines indicate the slopes of the outer $PCR$ curves}\label{fig:2}
\end{figure}

\quad For comparison with experiment, we have chosen the $PCR$ vs $I_B$ curves of WSi SNSPDs measured at different excitation wavelengths and bath temperatures \cite{Marsili1}
shown in Fig.\ref{fig:2}. To simulate these data we used the hotspot dynamics model developed in \cite{AGK}, modified to include Fano fluctuations.
In the absence of Fano fluctuations and spatial inhomogeneities this model predicts step-like $PCR$ curves.

\quad Fig.\ref{fig:2}a  shows the experimental and fitted $PCR$ vs $I_B$ curves at different temperatures. The fitting parameters were the same as in Marsili et al\cite{Marsili1} except for a higher photon yield $\bar{\chi}=0.32$, allowing the switching current to be $\sim60\%$ of the de-pairing  current. In the absence of measurements of the de-pairing current, its value is an additional  fitting parameter, allowing a better match between theory and experiment. For evaluation of $\sigma_2$ we used $J=1.3$. The calculated values of $J^1$ for WSi on a-SiO$_2$  range between 0.5 to 1.5 (for $\eta=0.5$) and  0.33 to 1.0 (for $\eta=0.3$) taking $\tau_0$ in the interval 2 to 10 ns. $l_{pe,D}$ varies in the limits from 0.8 nm for $\tau_0=$2 ns to 4 nm for $\tau_0=$10 ns.  Note that there is no information in the literature about the strength of elastic scattering of phonons in amorphous WSi film. For comparison, in a 6.9 nm-thick amorphous SiO$_2$ film, the phonon mean free path defining heat conductivity at $T>50$ K is comparable to film thickness\cite{Pohl}. The qualitative agreement between theory and experiment in Fig.\ref{fig:2}a is good. The simulated curves for the whole range of bath temperatures from 0.25 to 2 K are close to experimental results,  simulations also reproduce the steeper slopes of the $PCR$ curves with increased bath temperature. This feature is clearly seen in the experimental data and is an extra consistency check of our kinetic model.

\quad Also consistent with experimental data is the group of simulated $PCR$ curves for different photon wavelengths at fixed bath temperature, which have almost identical slopes. We used the same set of fitting parameters as in Fig.\ref{fig:2}a, fixing the bath temperature and allowing $\lambda$ to change. Steeper slopes at smaller bias currents (higher $T_s$ in fixed-$\lambda$ experiment) originate in the nonlinear $E(I_B,T_s,B)-E(I_B,T_b,B)$ dependence. The same nonlinearities are effective for similar range of currents determining the shape of $PCR$ curves also for the second experiment, i.e. $PCR$ vs bias current at fixed bath temperature and variable $\lambda$. However, in the second experiment, shown in Fig. \ref{fig:2}b, the effect of nonlinearity is nearly balanced by the  square root  dependence of variance $\sigma$ on photon energy. As a result the corresponding slopes remain almost unchanged, as expected from the model.

\subsection{$P_{click}$ vs $t_D$ curves}

\begin{figure}
  \centering
  \includegraphics[width=6cm]{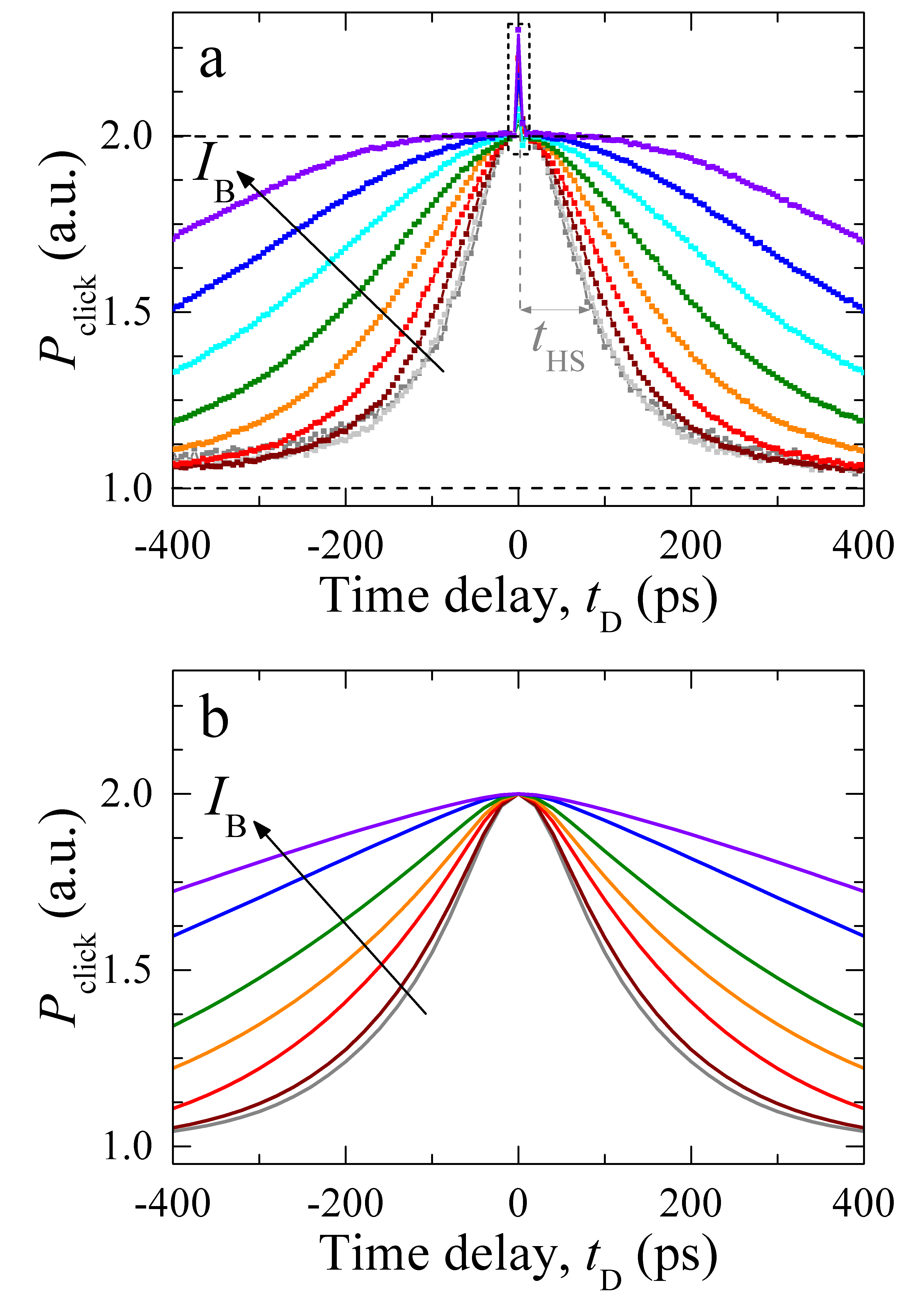}\\
  \caption{$PCR$ as a function of time delay $t_D$ for a series of bias currents: (a) experiment, (b) theory.}\label{fig:6}
\end{figure}

\quad We have shown that Fano fluctuations play an important role in smearing the detection threshold. This conclusion holds regardless of the ultimate detection mechanism, which primarily affects the magnitude of the
threshold energy $E^*$ and correspondingly the effective $T_s$. Here we discuss the role of Fano fluctuations in determining the hotspot relaxation dynamics. As reported in Ref.\cite{Marsili1}, we coupled optical pulse pairs separated by a variable delay $t_D$ to an SNSPD operating in the two photon regime. In the two photon detection regime, a photoresponse pulse can be efficiently triggered only if two photons generate two overlapping hotspots. As shown in Fig.\ref{fig:6}, we measured the probability of a response pulse, or click ($P_{click}$), as a function of $t_D$ for different bias currents. The $P_{click}$ vs $t_D$ curves have Lorentzian shape and become wider if the bias current is increased. We defined the hotspot relaxation time as the half width at half maximum of each Lorentzian curve.
This profile is evidenced by the shape of the two-photon $PCR(t_D)$ as a function of time delay between two single-photon pulses\cite{Marsili1}.

\quad In an idealized model with a sharp threshold for single-photon detection, $P_{click}$ vs $t_D$ curves will have a rectangular shape with a width determined by the hotspot relaxation time\cite{AGK}. According to our model, Fano fluctuations smear the sides of these rectangular curves, transforming them to bell-shape curves. Neglecting diffusive expansion\cite{AGK}, we let the first pulse at $t=0$ deposit an energy $E$  with probability $P(E)$, which creates a hotspot
with initial temperature $T_{ex}=T(E,T_b,I_B)$. Subsequently the hotspot relaxes, and its temperature follows the functional dependence $T(E,T_b,I_B,t)$. The latter can be found as a solution of the kinetic equation\cite{AGK}. At an instance
of time $t = t_D$ , the hotspot energy, $E_{HS}$, thus decreases
to $E_{HS}(T(E), T_b, I_B, t_D)$. Then the normalized two-photon $P_{click}$ becomes
\begin{eqnarray}\label{two-photon PCR}
PCR(t_D)=\int_0^\infty\mathrm{d}EP(E)\int_{E^*-E(T(E),T_b,I_B,t_D)}^\infty\mathrm{d}E'P(E')=\nonumber\\
\frac{1}{2}\int_0^\infty\mathrm{d}EP(E)\mathrm{erfc}\left[\frac{E(I_B,T_s)-E(T(E),T_b,I_B,t_D)-\bar{E}}{\sqrt{2}\sigma}\right]\nonumber\\~~~~~~~~~~~~~~~~~~~~~~~~~~~~~~~~~~~~(12)\nonumber
\end{eqnarray}
Fig.\ref{fig:6}b shows the simulated shapes of $P_{click}(t_D)$ for several different values of bias current using
the following parameters, $J=1.3$, $\bar{\chi}=0.32$, $I_{sw}/I_{dep}$=0.68  and $\tau_0=5.0$ ns.

\begin{figure}
  \centering
  \includegraphics[width=8 cm]{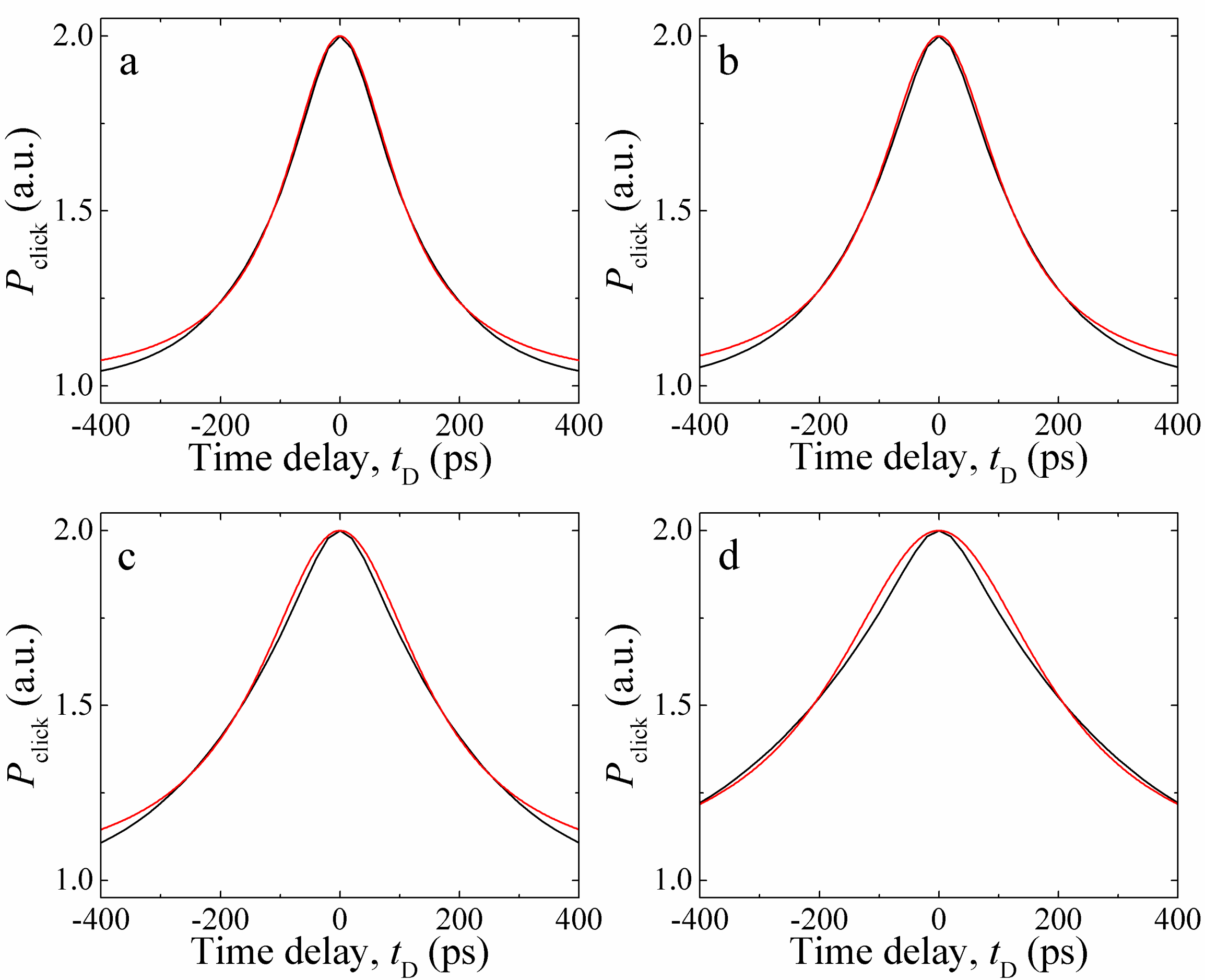}\\
  \caption{Fitting the simulated  $PCR(t_D)$ for series of  bias currents by the Lorentzians.}\label{fig:8}
\end{figure}

\quad As expected, Fano fluctuations play a dominant role in shaping the photoresponse $P_{click}(t_D)$, also providing further evidence that in amorphous WSi nanowires hotspot relaxation proceeds via self-recombination. While the simulated curves in Fig.\ref{fig:6} can be better matched to the experimentally observed Lorentzians over the hotspot lifetime defining range of time delays, their tails differ from experiment. The results of such a match are shown in Fig.\ref{fig:8}.
The simulated tails fall between that of Lorentzian and Gaussian. It is possible that diffusion makes most of its contribution at the tails of $PCR(t_D)$, causing reduced self-recombination at the periphery of the hotspot. Simulations in Fig.\ref{fig:6} look reasonably close to experimental curves.  They were done at parameter values close to our of earlier work\cite{AGK,Marsili1} assuming switching current $32\%$ less than de-paring current and correspondingly more efficient energy deposition parameter. In general larger difference between the switching and de-paring currents requires larger factors $\bar{\chi}$ to match theory and experiment. Physically that means that achieving the higher critical temperature in a wire with bias current being a smaller fraction of the de-paring current requires more deposited energy. Also,  decreasing $I_{sw}/I_{dep}$ and increasing $\bar{\chi}$ results in a weaker dependence of  hotspot relaxation time (half width at half maximum) on bias current as evidenced by a weaker dependence of hotspot relaxation time on $I_B$ at lower currents. This dependence becomes weaker also for larger values of $\tau_0$. Allowing $I_{sw}/I_{dep}$ to be free fitting parameter we may shift simulated $PCR$ curves along the bias current axis achieving better fit to experiment. These simulations must be considered together with best fits in Fig.\ref{fig:2} which required the same set of fitting parameters as for simulations in Fig.\ref{fig:6} except for $I_{sw}/I_{dep}$=0.6.  Achieving good fit with the same material parameters and fixed values for $I_{sw}/I_{dep}$, $\sigma$  and $\bar{\chi}$ is very challenging. We could have fitted each separate experiment better allowing some flexibility of each individual parameter. The fact that  we can semi-quantitatively fit all the different experiments,  allowing $12\%$ variation of a single $I_{sw}/I_{dep}$ fitting parameter and keeping the rest the same,  shows both the validity of the model and the accuracy of the main assumptions.

\quad Simulation of $PCR(t_D)$ for two-photon detection with variable time delay requires the two additional parameters, $\tau_0$, determining the time scale, and cutoff temperature, determining the level of cooling of hotspot below which it cannot be detected at arrival of the second photon. Both parameters are irrelevant for single photon detection. It is  important to note that for a given photon wavelength the ratio of cutoff to critical temperature $T_s$ is not an entirely independent parameter. It depends on the ratio of bias- to de-paring currents. However, this dependence is highly non-linear and its effect on simulations differs from the linear scaling of the bias relative to de-pairing current, which is sufficient for fitting single photon counting rates. In two-photon experiments with variable time delay, both linear re-scaling of the bias relative to depairing current and nonlinearity of ratio of $T_{co}$ to $T_s$  are important, and hence credibility of both sets of simulations depends on assumed applicability of BCS model. Finally,  the role of diffusion, which has not been discussed so far is also important. The simple model\cite{AGK,Marsili1}, although acknowledging its potential role, ignored it with some support from experimental data.

\subsection{Discussion}

\quad Our simulations of single-photon $PCR$-curves and two-photon pump-and-probe experiments with variable time delay were based on a narrow nanowire hot-belt model.
The analogous hot-belt model was claimed to be irrelevant for both NbN and WSi SNSPDs with nanowire width exceeding 150 nm\cite{Vodolazov4}.  Our experiments and simulations therefore allow testing of the hot-belt model predictions for a wide range of experiments and comparison to predictions of the hotspot model where we can identify expected differences. The hot belt model is expected to work for the two-photon detection experiment. It was demonstrated to produce qualitatively similar but quantitatively strongly different results to the hotspot model (except for SNSPDs in external magnetic fields in a certain range of bias currents). In contrast, the agreement with both experiments that comes as the result of our analysis is much closer than what could have been inferred on assumption that hot-belt and hotspot models work for two-photon and one-photon experiments respectively.

\quad The validity criterion for the narrow-wire model is $\tau_{th}\geq\tau_D$, where $\tau_{th}$  is the characteristic thermalisation time, controlling suppression of the gap within the initial volume, and $\tau_D$ is characteristic diffusion time across the nanowire\cite{Vodolazov4}.  We do not use the coherence length as the radius $R_0$ of phonon bubble. This scale is not relevant for the initial state, because the radius of phonon bubble  must be exactly the  same for a normal metal. We chose $R_0=\sqrt{4Dt_{d}+l^2_{pe,D}}$  as the more appropriate spacial scale. Here $t_d$ stands for the descent time for photoelectron(hole) from the level $E_1^*$ to Fermi energy. The meaning of $R_0$ is the length of the random walk, that the  primary photoelectron and hole perform while disposing their excess energy to phonons, creating a phonon bubble. The second term under the square root accounts for extra volume expansion due to phonons of the bubble moving on average a distance of their mean free path prior to being re-absorbed by electrons. For WSi using $d=5$ nm, $D=0.75 \mathrm{cm}^2/s$\cite{Marsili1}, $\tau_0=10$ ns, $\Omega_D=34$ meV\cite{sidorova1}, $\displaystyle{t_d=\tau_0\frac{2E_1^*}{\Omega_D}\left(\frac{T_c}{\Omega_D}\right)^3}$, $\displaystyle{\tau_{pe,D}=\frac{\tau_0}{\gamma}\frac{T_c}{\Omega_D}}$ and $\gamma$=89 we obtain for volume of phonon bubble $V_{0}=\pi£R_0^2d$=2044 nm$^3$ exceeding initial volume used for the estimate of $\tau_{th}$ in ref.\cite{Vodolazov4}. Correspondingly, the energy density is smaller resulting in lower temperature of of electrons and phonons in the hotspot $T_e=T_{ph}=4.1T_c$ and $\tau_{th}$=8.7 ps for absorption of a 1550 nm photon as  estimated from energy conservation. There is currently no consensus in the literature regarding the magnitude of $\tau_0$ with recent measurements of magnetoresistance\cite{sidorova2} yielding $\tau_0$=1.9 ns.  We have chosen $\tau_0$=10 ns for two reasons. It fits the measurements of electron-phonon relaxation time\cite{sidorova1} over the low temperature range. It also better fits the expected magnitude of $\tau_0$ inferred from scaling according to $\Omega_D^2/Tc^3$ law, which must work during the formation of phonon bubble when effects of disorder on electron-phonon interactions are not important. Phonon escape from the SNSPD film and diffusive expansion  during thermalisation both reduce the energy density within the volume filled with non-equilibrium excitations  resulting in further slowdown of thermalization. A more important process is likely to be diffusive expansion, which may result in a substantial increase of $\tau_{th}$. Indeed, during the first picosecond evolving hotspot with the initial volume of 2044 nm$^3$ expands to fill the volume of 4700 nm$^3$. Reducing energy density within the evolving hotspot results in the temperature of thermalised quasiparticles and phonons of $3.1T_c$ and $\tau_{th}>19.0$ ps ($>$9.5 ps if to assume $\tau_0$=5 ns). The representative value for  expansion time of hotspot across the width of the wire is the diffusion time from the centre of the strip, $\tau_D\sim£W^2/16D\simeq14.1$ ps. The account of both phonon loss and diffusion during the stage of thermalisation must be done within the refined model. In the absence of such a model and in view of significant uncertainty of material parameters the question of validity of one or the other model remains open.

\begin{figure}
  \centering
  \includegraphics[width=5cm]{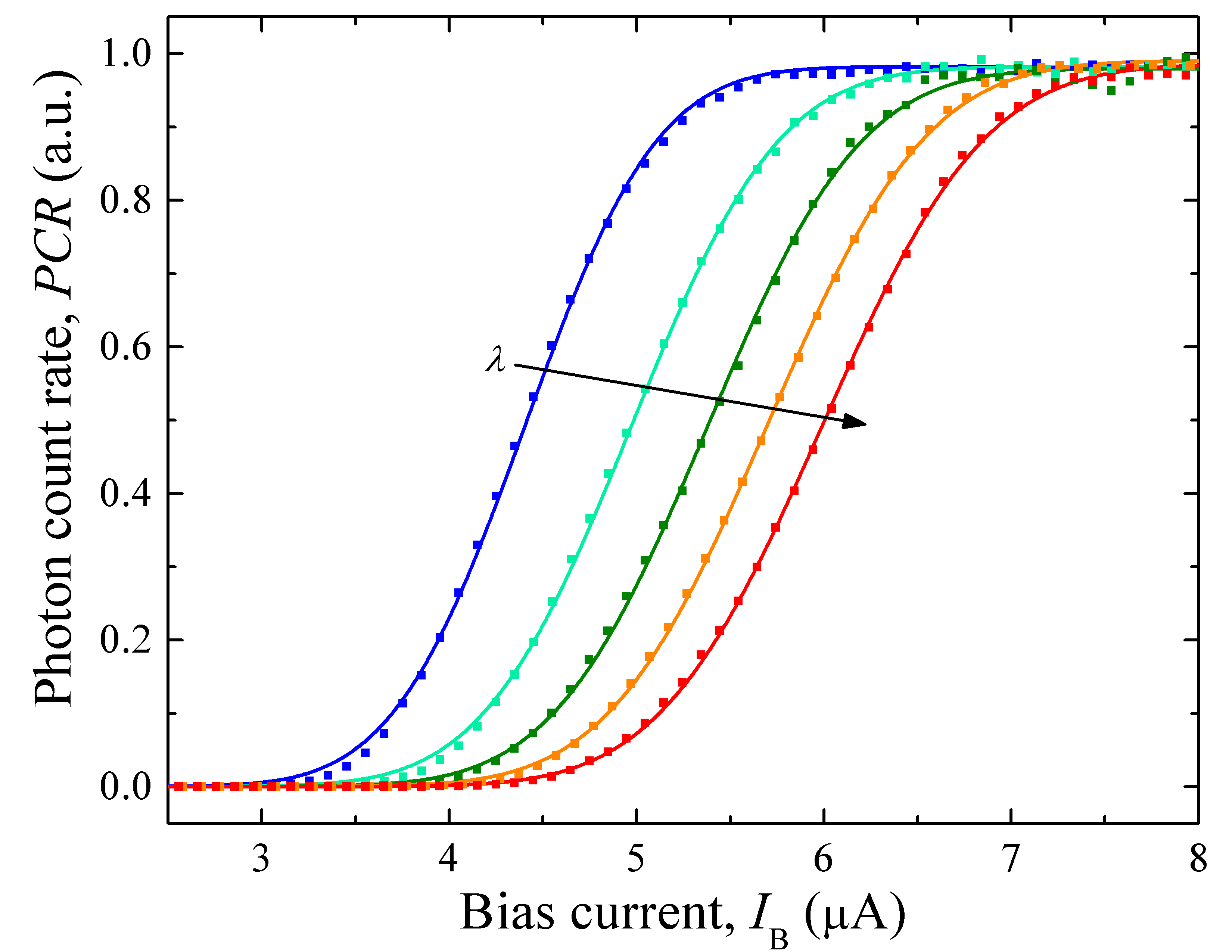}\\
  \caption{Fitting $PCR$ vs $I_B$  for $\lambda$ = 1200, 1350, 1450, 1550 and 1650 nm by error functions. }\label{fig:7}
\end{figure}

\quad The experimental data in Figs.\ref{fig:2}a and b have smooth sigmoidal shapes close to predicted error functions. Indeed, there is no evidence of coordinate-dependent response in the transition region. With any of the bell or w-shaped coordinate-dependent responses, one expects the specific change of curvature at the inflection point from concave below, $I_B<I_{co}$, to convex above cutoff current, $I_B>I_{co}$ as shown in Fig.\ref{fig:5}. In contrast, for a narrow wire, the predicted error function shape is convex on the left and concave on the right of the inflection point, $I_{co}$. In order to check whether the experimental curves in Fig.\ref{fig:2} can be approximated by error functions in Fig.\ref{fig:7} we show the results of matching the curves from Fig.\ref{fig:2}  to error functions, $\displaystyle{PCR=\frac{A}{2}\mathrm{erfc}\left(\frac{I_{co}-I_B}{\Delta£I}\right)}$, where ${A}$ and ${\Delta£I}$ are fitting parameters. No clear signs of coordinate-dependent response are seen in Fig.\ref{fig:7}.

\quad In a recent paper on MoSi SNSPDs, the $PCR$ curves were measured and fitted with error functions for a wavelength range of 750 to 2050 nm. It was found that  at low photon energies  the fit agrees very well with the data. However, at high energies, the shape of the curves starts to deviate from simple error function\cite{Caloz}. The authors hypothesize that these deviations might indicate a coordinate-dependent response of the SNSPD. The observed deviations, however, are not what can be expected from coordinate-dependent response. Firstly, the curvatures of $PCR$-shapes on both sides of inflection points remain consistent with fluctuation-induced homogeneous sensor response. Secondly, the experimental curve shows smaller counts both in the lower bias tail and on approach to saturation level. This is not indicative of an inhomogenious response, for which on approach to $I_{det}^{max}$ photon counting rate must exhibit a discontinuity in its derivative. It is possible that the observed deviations are connected with non-linearity of SNSPD response as given by the function $\alpha$ in expression (\ref{PCRIB}). This non-linearity is of the same origin as the observed substantial non-linearity of current-energy relation in MoSi\cite{Caloz}. These results may be also explained within the hotspot model provided that detection current is only weakly dependent on coordinates of absorption site so that predictions of hotspot model for the shape of $PCR$ vs $I_B$ will be no different and both models further merge as explained in Section II.

\quad The effects of an external magnetic field on $PCR$ vs bias current\cite{Bulaevskii2,Vodolazov3,Korneev,JJR} deserve special attention, since they can give insight into the detection mechanisms in SNSPDs\cite{Vodolazov3,Korneev}. Features due to Fano fluctuations were not accounted for in any of the previous work. There are several ways that external magnetic fields can combine with Fano fluctuations and thus cause the $PCR$ curves to change shape. In an external perpendicular magnetic field, the threshold $E^*$ depends on the field magnitude, as seen in (\ref{threshold}) and (\ref{PCR}). Both the critical current and the de-pairing energy depend on magnetic field, with the former being dominant for weak magnetic fields. With $I_c(B)$ decreasing, while $B$ increases, the obvious effect in weak fields is the shift of $PCR$ curves towards lower currents, as observed in \cite{Vodolazov3,Korneev,JJR}. The magnitude of the shifts depends on the exact functional dependence of $I_c(B)$.

\quad Vodolazov et al\cite{Vodolazov3,Korneev}  observed shape transformations of the $PCR$ curves with varying magnetic field and photon wavelength above the crossover current. Their results were interpreted on the basis of a model assuming vortex trapping by compact hotspots having a radius depending on the photon wavelength. The SNSPD response in a weak magnetic field was suggested as the definitive experiment for identification of the detection mechanism\cite{Vodolazov4}.  Within our model with uniform current density, there is no such a crossover.  Nonetheless, the shifts may assume a more complicated pattern due to the interplay between the critical current, the de-pairing energy, and the complicated nonlinear $E(I_B,T_s,B)-E(I_B,T_b,B)$ dependence. At shorter photon wavelengths, the $PCR$ curves shift towards lower bias, where the contribution of magnetic field term to depairing energy increases relative to the supercurrent term. Significant non-linearity of the energy-current relation was recently reported over the spectral interval 750-2050 nm\cite{Caloz}. A full understanding of the detection mechanisms in SNSPDs will require a more detailed study of $PCR$ shapes through the transition range at the low current tail, near inflection point(s) and close to saturation together with thorough study of the role of magnetic field affecting  non-linearities of response and Fano effect.

\section{Conclusions}
\quad In summary, we have shown that Fano fluctuations play a fundamental role in superconducting nanowire single-photon detectors. They are an essential factor in determining the exact shape of both single-photon photoresponse and time-delayed, two-photon photoresponse. The special features of $PCR(I_B)$ curves, such as slope transformation, positions of inflection point(s) versus wavelength, bath temperature and magnetic field reveal a wealth of sensor physics and will do a significant service for unambiguous determination of the detection mechanism(s).

\section{Acknowledgements}

\quad AGK and CL acknowledge financial support from the Engineering and Physical Sciences Research Council.
AGK, FM and MDS acknowledge financial support from DARPA, J.P.A. was supported by a NASA Space Technology Research Fellowship.

\onecolumngrid

 \end{document}